\documentclass[12pt,submission,journal,onecolumn]{IEEEtran}
\ifCLASSINFOpdf
\else
\fi

\usepackage{epsfig,amsmath,amssymb,amsthm}
\usepackage{graphicx}
\usepackage{cite}
\usepackage{times}
\usepackage{algorithm}
\usepackage{array}
\usepackage{algorithmic}
\usepackage{url}
\usepackage{indentfirst}
\usepackage{subfigure}
\usepackage{multicol}
\usepackage{amsfonts}
\usepackage{fancybox}
\usepackage{enumerate}
\usepackage{psfrag}
\usepackage[margin=0.8in]{geometry}

\begin{document}
\title{On the Feasibility of Indoor Broadband Secondary Access to 960-1215 MHz Aeronautical Spectrum}

\author{Evanny Obregon,~Ki Won Sung, and~Jens Zander
\thanks{E. Obregon, K. W. Sung and J. Zander are with KTH Royal Institute of Technology, Wireless@KTH, SE 164 40, Stockholm, Sweden (e-mail: ecog@kth.se, sungkw@kth.se, jenz@kth.se)}
}
\mark{Submitted for Publication}
\maketitle
\begin{abstract}
In this paper, we analyze the feasibility of indoor broadband service provisioning using secondary spectrum access to the 960-1215 MHz band, primarily allocated to the distance measuring equipment (DME) system for aeronautical navigation. We propose a practical secondary sharing scheme customized to the characteristics of the DME. Since the primary system performs a safety-of-life functionality, protection from harmful interference becomes extremely critical. The proposed scheme controls aggregate interference by imposing an individual interference threshold on the secondary users. We examine the feasibility of large scale secondary access in terms of the transmission probability of the secondary users that keeps the probability of harmful interference below a given limit. Uncertainties in the estimation of propagation loss and DME location affect the feasibility of the secondary access. Numerical results show that large number of secondary users are able to operate in adjacent DME channels without harming the primary system even with limited accuracy in the estimation of the propagation loss.
\end{abstract}


\section{Introduction}
\label{sec:Introduction}

It is generally believed that spectrum shortage is caused by inefficient spectrum utilization under the existing regulatory and licensing process that only allows static spectrum allocation. Secondary spectrum access has emerged as a promising solution to relieve the apparent spectrum shortage~\cite{hskz1201}. In spite of extensive theoretical research on the field of cognitive radio and dynamic spectrum access, the practical value of the secondary access has not been fully investigated. Most of efforts to assess the real-life benefit of the secondary spectrum have thus far focused on the digital TV broadcasting bands, namely TV white spaces~\cite{hms1001,bram1201}. Substantial portion of useful spectrum is primarily allocated to various systems such as radar and aeronautical navigation, whose secondary access feasibility is mostly unexplored~\cite{hskz1201}.

This work focuses on the 960-1215 MHz band which is allocated to aeronautical systems. In particular, this frequency band is mainly occupied by distance measuring equipment (DME). Secondary access to the 960-1215 MHz band was first studied in our previous work~\cite{5936226}. As a first step, we investigated the minimum requirements for the secondary users under the ideal assumption that the secondary users have accurate knowledge of propagation loss to the DME receivers. We observed that the secondary usage would be widely available under this particular assumption. However, it is obvious that the requirements to the secondary users will become more stringent if there are uncertainties in the propagation information. In practice, it is difficult to have perfect knowledge of the propagation to the DME system. Thus, it is needed to study the feasibility of secondary access under practical assumptions.

In this paper, we investigate the feasibility of secondary spectrum sharing with the DME system. To our best knowledge, it is the first attempt to examine the practical usefulness of 960-1215 MHz with regard to the secondary access. Our contribution can be detailed as follows. First, we propose practical methods by which the secondary users discover opportunities and share the spectrum. They are customized to the characteristics of the primary user, i.e. DME receivers, and based on geo-location database and spectrum sensing. Second, we identify the major sources of uncertainties that cause inaccurate estimation of propagation loss to the DME, and analyze the impact of the uncertainties by employing mathematical aggregate interference models in~\cite{gs0801,5701700}.

We consider massive deployment of secondary users that provide high-speed indoor broadband, e.g. WiFi and HeNB. Such a large scale secondary access is deemed \emph{feasible} if the practical sharing methods enable the secondary users to maintain an acceptable transmission probability. Since our analysis accompanies the uncertainties in the propagation loss estimation, we focus on the following research questions:

\begin{itemize}
\item Is the massive secondary access feasible in 960-1215 MHz band?
\item What is the impact of the uncertainties on the feasibility of secondary access?
\end{itemize}

The rest of the paper is organized as follows: the system model, primary and secondary systems characteristics are described in Section~\ref{sec:SystemModel}. The proposed secondary access scheme and the mathematical models of the aggregate interference for ground transponder and airborne interrogator are introduced in Section~\ref{sec:interferenceGround} and in Section~\ref{sec:interferenceAirborne}, respectively. In Section~\ref{sec:NumericalResults}, we present and discuss our numerical results. Finally, main conclusions of this work and remaining issues for future studies are given in Section~\ref{sec:Conclusions}.

\section{System Model}
\label{sec:SystemModel}

\subsection{DME as the primary system}
\label{sec:DME}
DME is used for measuring the distance between an aircraft and a ground station. The airborne equipment (interrogator) sends short Gaussian pulses down to earth, and the ground station (transponder) responds on a frequency of $\pm$63~MHz from the interrogation frequency. The interrogator can calculate the slant distance based on the round trip delay of the signal. The pulses are burst more than 100 times per second by the interrogator and 2500 times by the transponder. Their transmission power reach up to 300~W for the interrogator and up to 2~kW for the transponder. The channel bandwidth of DME is 1~MHz, i.e. there are 252 channels in total. More detailed operation of DME can be found in~\cite{5936226} and references therein.

We consider that the DME receiver can tolerate a maximum interference power of $A_{thr}$, which corresponds to $-119$dBm/MHz and $-111$dBm/MHz for the transponder and interrogator, respectively~\cite{5936226}. The received interference is considered harmful if it exceeds $A_{thr}$.
Because the DME system performs a safety-of-life functionality, protection from harmful interference becomes extremely critical. Due to the high sensitivity of DME receivers, we need to control the aggregate interference over a large area, which is the major challenge for the secondary access to this spectrum. The aggregate interference ($I_{a}$) is regulated as follows:
\begin{equation}
\label{eq:ProtectionDME}
\Pr[I_{a}> A_{thr}]\leq\beta_{PU}
\end{equation}
where $\beta_{PU}$ is the maximum permissible probability of harmful interference at the primary receiver. The nature of DME operation requires $\beta_{PU}$ to be extremely small. A reasonable range of $\beta_{PU}$ has not been discussed well in the literature. We adopt a value used for air traffic control radar in 2.7-2.9 GHz, i.e. $\beta_{PU} = 0.001\%$~\cite{ImadurPIRMC}.

The interference from the DME device to the secondary receiver is, on the contrary, negligible, since the DME generates only short pulses. Although the DME pair exchanges the pulses frequently, the overall channel utilization remains below 1\%. Secondary receivers might be saturated if they receive excessively strong DME pulses.  Let $I_{sat}$ be the saturation point of the secondary receiver. Then, the following condition should be satisfied:
\begin{equation}
\label{eq:ProtectionSU}
\Pr[I_{PU}>I_{sat}]\leq\beta_{SU}
\end{equation}
where $\beta_{SU}$ is the maximum saturation probability and $I_{PU}$ is the received primary pulse power. We adopt a value of $\beta_{SU} = 2\%$ and $I_{sat} = -30$dBm which is a typical saturation level of low noise amplifier (LNA) in WiFi receivers \cite{wp:ti2003}. With the adopted values for $A_{thr}$, a simple link budget analysis indicates that \eqref{eq:ProtectionDME} is the limiting constraint even before taking the effect of multiple secondary users into account. Therefore, we will focus on the protection of the primary user in the remainder of the paper.

\subsection{Indoor Broadband as secondary system}

Let us consider massive scale deployment of indoor access points and mobiles for high capacity broadband services over a large area. They utilize the spectrum allocated to the DME under the principle of spectrum interweave~\cite{4840529}. The secondary users are assumed to be spatially distributed according to a homogeneous Poisson point process in a two dimensional plane $\Re^2$. The primary receiver is located at the center of the circular region limited by two radii $r_{o}$ and $R$, which are the minimum and maximum distances from the primary receiver, respectively.

Each secondary user decides whether it can access a particular DME channel or not by estimating the interference it will generate to the primary user. Let $I_{thr}$ denote the interference threshold imposed on the individual secondary users. The value of $I_{thr}$ is given to the secondary users by a central spectrum manager. This ensures that each secondary users makes its own decision without interacting with the others. The interference from a secondary user $i$ is given by

\begin{equation}
\label{eq:InterferenceRule}
I_{i}= \left\{ \begin{array} {rl} \xi_{i}, &\mbox{ if $\tilde{\xi}_{i}\leq I_{thr}$}\\
0, &\mbox{ otherwise}
       \end{array} \right.
\end{equation}
where $ \xi_{i}$ is the interference that the primary user would receive if an arbitrary secondary user were to transmit, and $\tilde{\xi}_{i}$ is the estimate of $ \xi_{i}$ by the secondary user $i$. Note that $\xi_{i} = \tilde{\xi}_{i}$ only when the secondary user has the perfect knowledge of the propagation loss. Considering that there are $N$ secondary users around the primary user, the aggregate interference is
\begin{equation}
\label{eq:Ia}
I_{a}= \sum_{i\in N_{t}} I_{i}
\end{equation}
where $N_{t}$ is the set of transmitting secondary users.

\section{Sharing with the ground transponder}
\label{sec:interferenceGround}

\subsection{Secondary access scheme}
\label{subsec:GroundScheme}
The ground transponder is placed at a fixed location and frequently bursts short pulses to the airborne interrogators. Thus, it is possible for the secondary user to detect the existence of the transponder via spectrum sensing. The additional use of geo-location database enables the secondary users to have prior knowledge about the transponder such as the location, operating frequency, and transmission power. This will significantly improve the performance of the spectrum sensing since the secondary users can have a good expectation about to signal to detect. Given the high transmission power of the transponder, we assume that the spectrum sensing is reliable enough to ignore missed detection and false alarm.

\begin{figure}[ht]
  \centering \includegraphics[width=.58\textwidth]{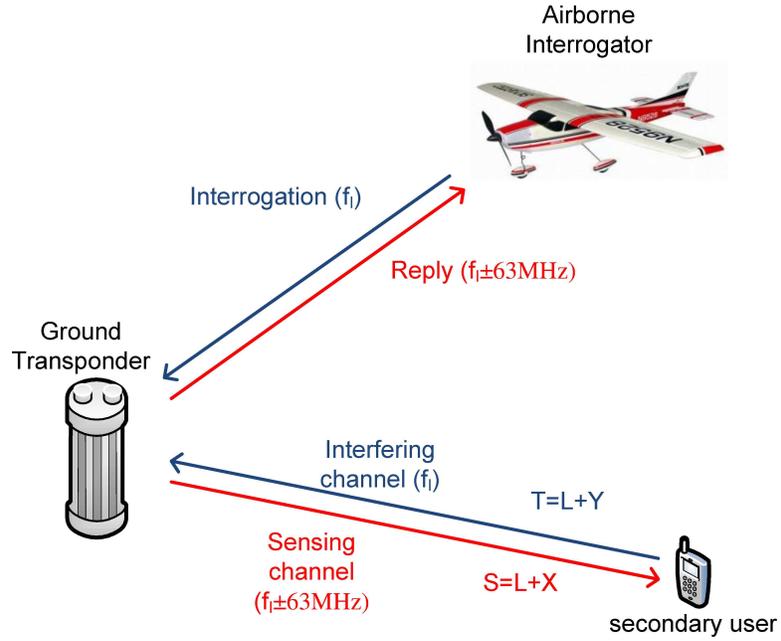}\\
  \centering \caption{Uncertainty in secondary sharing scenario with ground transponder as primary victim}
  \label{fig:GroundScenario}
\end{figure}

Fig.~\ref{fig:GroundScenario} depicts the proposed opportunity detection mechanism. Notice that the secondary users detect the transponder on the reply (sensing) frequency, while the interference is given on the interrogation (interfering) frequency. In both channels, propagation losses between the DME transponder and the secondary user consist of the distance-based path loss ($L$) and fading\footnote{Note that the fading here refers to the combined effect of shadowing and multi-path fading} ($X$ and $Y$). Although it is reasonable to assume that the secondary users accurately estimate the propagation loss of sensing channel ($S=L+X$), it does not necessarily mean that the estimation of interfering channel ($T=L+Y$) is also accurate. With the frequency offset of 63 MHz between the sensing and interfering channels, the shadowing components are typically highly correlated $(\rho_{shadowing}\approx1)$\cite{188579}, while the multi-path fading is uncorrelated $(\rho_{fast}=0)$. Therefore, the correlation between the composite fading components, $\rho$, lies between $[0,1]$. The exact value of $\rho$ depends on the characteristics of different propagation environments. Partial correlation between channels does not allow the secondary user to perfectly estimate its interference to the primary victim. Then, an \emph{uncertainty} in the estimation of fading component of the propagation loss between the secondary user and the ground transponder still remains. 

\subsection{Aggregate interference modeling}
\label{subsec:interferenceGround}
In this section, we model the aggregate interference when there is uncertainty in the fading estimation. Different levels of uncertainty in fading estimation are represented by a correlation coefficient $\rho$. We adopt the mathematical frameworks proposed in \cite{5936226,gs0801,5701700} with a slight modification to account for the proposed spectrum sharing mechanism.

Let us consider an arbitrary secondary user $i$ which is distributed according to a homogeneous Poisson point process in a circular area of radius $R$. The path loss between the primary receiver and the secondary user $i$ is modeled as $g(r_{i})=Cr_{i}^{-\alpha}$ where $C$ is a constant and $\alpha$ is the path loss exponent. Then, the user $i$ would cause interference $\xi_{i}$ to the primary receiver if it were to transmit, which can be expressed as
\begin{equation}
\label{eq:RealInter}
\xi_{i}=P_{t}^{eff} g(r_{i}) Y_{i}
\end{equation}
where $P_{t}^{eff}$ refers to the effective transmission power of the secondary user including antenna gains and bandwidth mismatch. $Y_{i}$ is a random variable modeling the fading effect. It is generally considered that the fading consists of shadow fading following a normal distribution in dB scale and multi-path fading by which the instantaneous power is varied with an exponential distribution. We use the assumption that the composite fading $Y_{i}$ follows a log-normal distribution. It is known that this assumption works well when the standard deviation of shadowing is higher than 6dB, i.e. when the shadowing is a dominant factor of the composite fading~\cite{507537}.

The user $i$ will decide to transmit if $\tilde{\xi}_{i}\leq I_{thr}$. Note that $\tilde{\xi}_{i}$ is affected by the fading on the sensing channel. That is,
\begin{equation}
\label{eq:BelievedInter}
\tilde{\xi}_{i} = P_{t}^{eff} g(r_{i}) X_{i}
\end{equation}
where $X_{i}$ is modeled as a log-normally distributed random variable whose parameters are same as $Y_{i}$. Therefore, the joint distribution of $X_{i}$ and $Y_{i}$ is given by the following bivariate log-normal distribution:
\begin{equation}
\label{eq:bivariatelog}
f_{X_{i},Y_{i}}(x,y) = \frac{1}{2 \pi xy  \sigma^2 \sqrt{1-\rho^2} } e^{-\frac{(\ln x)^2-2\rho(\ln x)(\ln y)+ (\ln y)^2}{2 \sigma^2 (1-\rho^2)}}
\end{equation}
where $\rho$ is the correlation coefficient of $X_{i}$ and $Y_{i}$:
\begin{equation}
\label{eq:correlation}
\rho = \frac{Cov(\ln X_{i},\ln Y_{i})}{\sqrt{Var(\ln X_{i})Var(\ln Y_{i})}}.
\end{equation}

We consider that the composite fading components, $X_{i}$ and $Y_{i}$, will be partially correlated ($ 0 < \rho < 1$). The exact value of $\rho$ depends on propagation environments. Note that full correlation ($\rho = 1$) represents an ideal case that the secondary user has an accurate knowledge of interference. On the opposite, zero correlation ($\rho = 0$) stands for a pessimistic assumption that the fading is completely unknown to the secondary user. For simplicity and mathematical tractability, we have adopted the assumption that secondary users in the whole area of study are affected by a homogeneous fading distribution. The feasibility of secondary access under different assumptions, ranging from ideal to pessimistic, will be shown and discussed in Section~\ref{sec:NumericalResults}.

The aggregate interference $I_{a}$ can be expressed as:
\begin{equation}
\label{eq:Ia2}
I_{a}= P_{t}^{eff} C \underbrace{\sum_{i\in N_{t}} r_{i}^{-\alpha} Y_{i}}_{I_{N_{t}}}.
\end{equation}
Hereafter, we omit the index of secondary user $i$, which is chosen in an arbitrary manner, unless necessary. By applying the Campbell's theorem, the characteristic function of $I_{N_{t}}$ is as follows:
\begin{equation}
\label{eq:characteristicfunction}
\begin{split}
\psi_{I_{N_{t}}}(jw) & = \exp \biggl (-2 \pi \lambda \int_{X} \int_{Y} \int_{r_{o}}^{R} [1-\exp(jwyr^{-\alpha})] \\
& \quad \times \textbf{1}_{[0,\hat{I_{thr}}]}(r^{-\alpha}x) f_{X,Y}(x,y) r\mathrm{d}r \mathrm{d}y \mathrm{d}x \biggl).
\end{split}
\end{equation}
where $j=\sqrt{-1}$ and $\hat{I_{thr}} = I_{thr}/(P_{t}^{eff}C)$. The activity of the secondary users is represented by $\textbf{1}_{[0,\hat{I_{thr}}]}(r^{-\alpha}x)$, which is a Bernoulli random variable. The indicator function is defined as:
\begin{equation}
\label{eq:indicator}
\textbf{1}_{[a,b]}(z)= \left\{ \begin{array} {rl} 1, &\mbox{ if $a\leq z\leq b$}\\
0, &\mbox{ otherwise}
       \end{array} \right.
\end{equation}
where the value one of the Bernoulli variable denotes that the secondary user is able to transmit. We use (\ref{eq:characteristicfunction}) to derive exact expressions for the $n^{th}$ cumulant of the aggregate interference in a limited circular region [$r_{o}, R$]. We consider the case where there is a partial correlation between the two fading effects affecting the sensing and interfering channels, $X$ and $Y$.
\begin{equation}
\label{eq:cumulatlog}
\begin{split}
k_{I_{N_{t}}} (n) & = \frac{2 \pi \lambda}{(n\alpha-2)} \biggl[(r_{o}^{2-n\alpha}-R^{2-n\alpha}) \int_0^\infty y^n f_{Y}(y) \Phi(L_{i})\mathrm{d}y\\
& \quad - R^{2-n\alpha} \int_0^\infty y^n f_{Y}(y)[\Phi(L_{s})-\Phi(L_{i})]\mathrm{d}y\\
& \quad +  \hat{I_{thr}}^{\frac{n\alpha-2}{\alpha}} \int_0^\infty y^n f_{Y}(y) \\
& \quad \times \int_{r_{o}^\alpha  \hat{I_{thr}}}^{R^\alpha  \hat{I_{thr}}} \frac{x^{\frac{2-n\alpha}{\alpha}}}{\sqrt{2\pi} x \sigma \sqrt{1-\rho^2}} e^{-\frac{(\ln x-\rho\ln y)^2}{2\sigma^2(1-\rho^2)}} \mathrm{d}x \mathrm{d}y\biggl]\\
\end{split}
\end{equation}
where,

\begin{equation*}
L_{i} = \frac{\ln(r_{o}^\alpha  \hat{I_{thr}})-\rho \ln y}{\sigma \sqrt{1-\rho^2}},
\end{equation*}
\begin{equation*}
L_{s} = \frac{\ln(R^\alpha  \hat{I_{thr}})-\rho \ln y}{\sigma \sqrt{1-\rho^2}}.
\end{equation*}

For the special cases of full correlation ($\rho = 1$) and zero correlation ($\rho = 0$), the closed-form expressions of cumulants can be found in \cite{gs0801} and \cite{5701700}, respectively. Using the cumulant of $I_{N_{t}}$ shown in (\ref{eq:cumulatlog}), we can obtain the $n^{th}$ cumulant of the aggregate interference $I_{a}$ as follows:
\begin{equation}
\label{eq:cumulantfinal}
k_{I_{a}} (n) = (P_{t}^{eff} C)^n k_{I_{N_{t}}} (n).
\end{equation}

The probability density function (pdf) of $I_{a}$ can be approximated with a known distribution by moment-matching method. In \cite{gs0801,5701700}, shifted log-normal and truncated-stable distributions are employed to address the skewness of the aggregate interference. In our model, the strong interferers are effectively removed due to the stringent threshold in~\eqref{eq:InterferenceRule}. Therefore, simple log-normal distribution sufficiently describes $I_{a}$. The pdf of $I_{a}$ can be approximated with the first and second order cumulants of $I_{a}$ obtained by \eqref{eq:cumulantfinal}.

\begin{equation}
\label{eq:ground_fIa}
f_{I_{a}} (y) = \frac{1}{y\sqrt{2\pi\sigma_{I_{a}}^2}} \exp\biggl[\frac{-\ln y -\mu_{I_{a}}}{2\sigma_{I_{a}}^2}\biggl],
\end{equation}
where
\begin{equation}
k_{I_{a}}(1) = \exp[\mu_{I_{a}}+\sigma_{I_{a}}^2/2],
\end{equation}
\begin{equation}
k_{I_{a}}(2) = \exp(\sigma_{I_{a}}^2-1)\exp(2\mu_{I_{a}}+\sigma_{I_{a}}^2).
\end{equation}

\section{Sharing with the airborne interrogator}
\label{sec:interferenceAirborne}

\subsection{Secondary access scheme}
\label{subsec:AirborneScheme}
Airborne interrogators are equipped in the airplanes, which are moving with a high speed. Therefore, it is not reasonable to assume a reliable detection of the interrogator via spectrum sensing. Instead, we assume that the secondary users are connected to a real-time database where the locations of the airplanes are provided. A living example of such a real-time aircraft location map can be found in \cite{misc:RDB}. Currently, the database information is updated every 20-60 seconds and has a limited coverage, which means that some airplanes (mostly small ones) do not appear in the map. However, we expect that an official database in a future will be able to provide a reliable information since it will be maintained by national authorities.

\begin{figure}[ht]
  \centering \includegraphics[width=.58\textwidth]{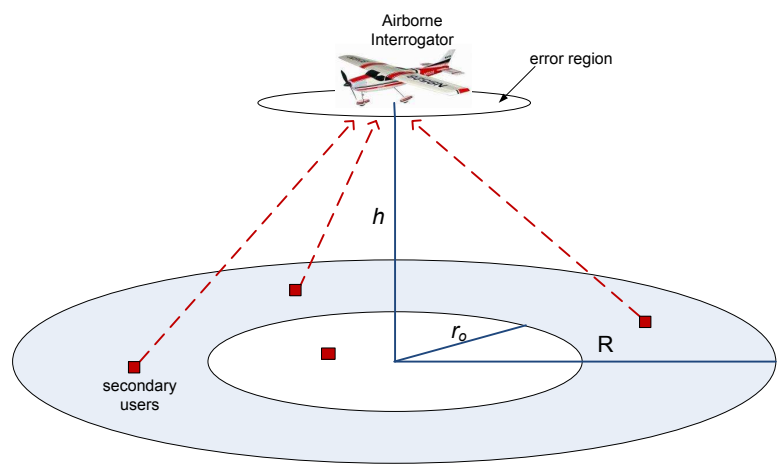}\\
  \centering \caption{Uncertainty in secondary sharing scenario with airborne interrogator as primary victim}
  \label{fig:AirborneScenario}
\end{figure}

Due to the \emph{update delay} in the database, the secondary user could potentially experience \emph{uncertainty} or imperfect information on the location of the airborne interrogator which is changing rapidly. Based on the update delay and the speed of the airplane, we introduce the notion of \emph{error region}, inside which secondary users will assume the worst case scenario that the sky is full of airplanes as shown in Fig. \ref{fig:AirborneScenario}. Outside the error region, secondary users will assume that the primary receiver is located at the closest border of the error region. Let $t_{u}$ be the time of update delay and $v$ be the speed of the airplane. Then, the radius of the error region is given by $t_{u} v$. For example, The $t_{u}$ of one minute corresponds to the error region of 15 km radius assuming $v=900$ km/h.

\subsection{Aggregate interference modeling}
\label{subsec:AirborneModel}
For the case of the airborne interrogator, free-space propagation model between the secondary users and the primary receiver is assumed. This means that fading effect is not taken into account. We adopt this assumption in order to account for the worst case scenario where there exists line-of-sight path between every secondary user and the primary user.

Similar to the ground transponder case, we assume that $N$ secondary users are distributed according to a homogeneous Poisson point process in a circular area of radius $R$. The primary victim is assumed to be located at the center with a height of $h$ from the ground. Since the fading effect is not considered, applying individual threshold $I_{thr}$ will result in a circular exclusion region where secondary users are not allowed to transmit. The radius of exclusion region is denoted by $r_o$.

Let $r_{thr}$ be the exclusion radius under the assumption that the secondary users know the exact location of the primary victim. In the presence of the update delay, each secondary user has to make a conservative decision that the airplane is at the closest possible location. It effectively increases the exclusion radius by $t_{u} v$. However, if the exclusion region is not needed in the first place ($r_{thr}=0$), the uncertainty in the primary user location does not make any impact on the feasibility of the secondary users. Thus, $r_o$ is given by
\begin{equation}\label{eq:AirborneRo}
   r_o= \left\{ \begin{array} {ll} r_{thr} + t_{u} v, &\mbox{if $r_{thr}>0$,}\\
0, &\mbox{otherwise.}
       \end{array} \right.
\end{equation}

Let $l_i$ be the distance from an arbitrary secondary user $i$ to the primary receiver. Then, $l_i = \sqrt{h^2+{r_i}^2}$ and the path loss $g(l_{i})$ is given by $C l_i^{-\alpha}$. Then, the aggregate interference $I_a$ is

\begin{equation} \label{eq:air_Ia}
I_a = P_{t}^{eff} C \underbrace{\sum^{N_t}_{i=1} l_{i}^{-\alpha}}_{I_{N_{t}}}
\end{equation}
where $N_t$ is the number of secondary users that are allowed to transmit, i.e. located at the outside of the exclusion region. Similar to Section~\ref{subsec:interferenceGround}, we apply the Campbell's theorem to obtain the characteristic function of $I_{N_{t}}$. Then, we derive exact expressions for the $n^{th}$ cumulant of $I_{N_{t}}$ in a limited circular region [$r_{o}, R$].
\begin{equation}
\label{eq:air_cumulant}
\begin{split}
k_{I_{N_{t}}} (n) & = \frac{2 \pi \lambda}{n\alpha - 2} (B^{(2-n\alpha)/2}-A^{(2-n\alpha)/2}).
\end{split}
\end{equation}
where $A=h^2+R^2 ~ \textrm{and} ~ B=h^2+r_o^2.$ Since we consider the free-space propagation model ($\alpha = 2$), we employ l'Hopital's rule to calculate the first order cumulant ($k_{I_{N_{t}}} (1)$).
Using the cumulant of $I_{N_{t}}$, we can obtain the $n^{th}$ cumulant of the aggregate interference $I_{a}$ as it is shown in~\eqref{eq:cumulantfinal}.
Note that $I_i$ is only affected by the distance-based path loss. Thus, $I_a$ is well described by the central limit theorem. This means that $I_a$ can be approximated as a Gaussian distribution with the first two cumulants as the mean and variance.

\section{Numerical Results}
\label{sec:NumericalResults}
The parameters used for our numerical experiments are described in Table \ref{tab:param}. For the case of ground transponder, we model the propagation loss between the primary victim and the secondary user using Hata model for suburban area. Instead, for airborne interrogator we employ free-space propagation loss. For the transponder, we investigate the impact of $\rho$ on the requirement and feasibility of secondary access in terms of the individual interference threshold $I_{thr}$ and the transmission probability of the secondary user $i$, $\Pr(\tilde{\xi}_{i}\leq I_{thr})$, at a given $r_{i}$. For the interrogator, we analyze the effect of the update delay on the requirements of secondary users. The feasibility of secondary access is given in terms of the exclusion region size $r_{o}$ imposed on the secondary users.

In both cases, we provide results for co-channel usage and as well as adjacent channel usage. We apply DME selectivity mask given in~\cite{ecc96} to determine the adjacent channel rejection (ACR) characteristics. This means that the condition~\eqref{eq:ProtectionDME} is changed to $\Pr[I_{a}> (A_{thr}+ ACR)]\leq\beta_{PU}$ when we evaluate the adjacent channel usage. The values of ACR will vary according the frequency separation. For instance, it is between 60dB and 70dB for channels with a frequency separation of 2 MHz. We assume that this applies as well to the channels of more frequency separation. To account for interference aggregation in the spectral domain, we apply a fixed margin of 3dB and 10dB for co-channel and adjacent channel, respectively.

\begin{figure}[h]
  \centering \includegraphics[width=.68\textwidth]{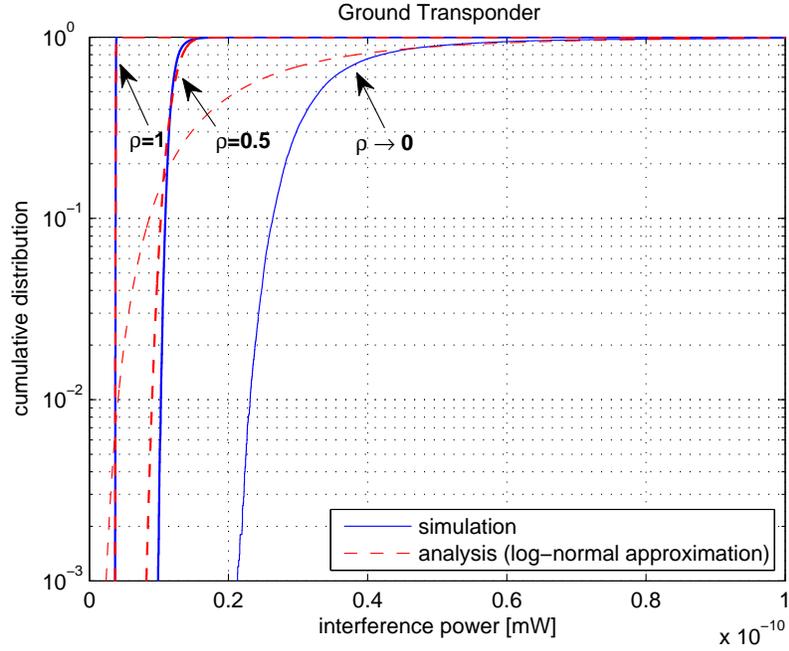}\\
  \centering \caption{A comparison between the analytic CDF of $I_{a}$ and the result of Monte Carlo simulation; primary receiver is the DME ground transponder ($I_{thr}=-150$dBm and $\lambda_{SU}=20/km^2$)}
 \label{fig:SimulationVsAnalysis}
\end{figure}

For the case of ground transponder, the cumulative distribution function (CDF) of $I_{a}$ calculated from (\ref{eq:ground_fIa}) for different values of $\rho$ is shown in Fig. \ref{fig:SimulationVsAnalysis}. A good agreement between analytical CDF of $I_{a}$ and the simulation results is verified when $\rho>0$. When the fading is unknown to the secondary user ($\rho = 0$), analytical CDF matches the tails of the simulation-based CDF of $I_{a}$. Since we are working with $\beta_{PU} = 0.001\%$, it is still possible to employ the log-normal approximation of the probability distribution of $I_{a}$ to analyze the impact of fading uncertainty on the feasibility of secondary access.

\begin{figure}[h]
  \centering \includegraphics[width=.68\textwidth]{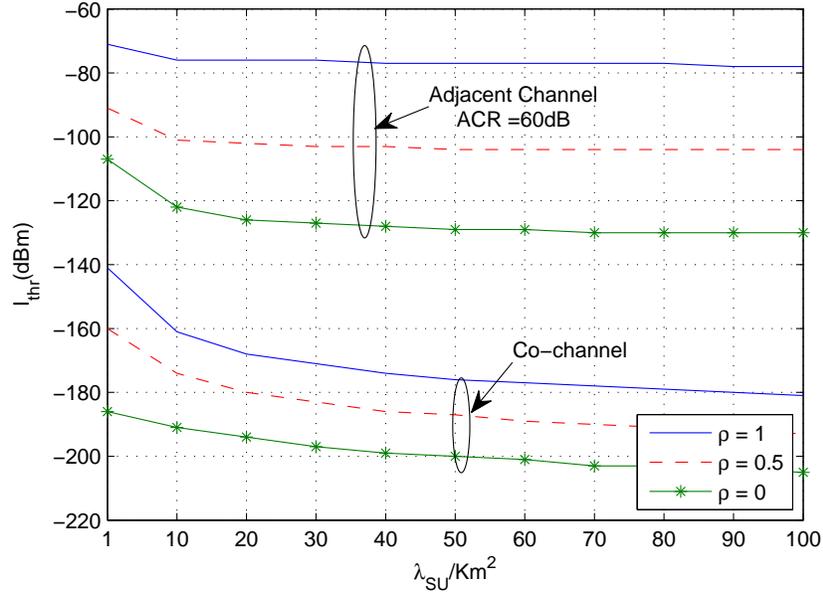}\\
  \centering \caption{$I_{thr}$ as a function of secondary network density ($\lambda_{SU}$) for different correlation coefficients($\rho$); the primary receiver is the DME ground transponder}
  \label{fig:IthrDiffCorrelation}
\end{figure}

\begin{figure}[h]
 \centering \includegraphics[width=.68\textwidth]{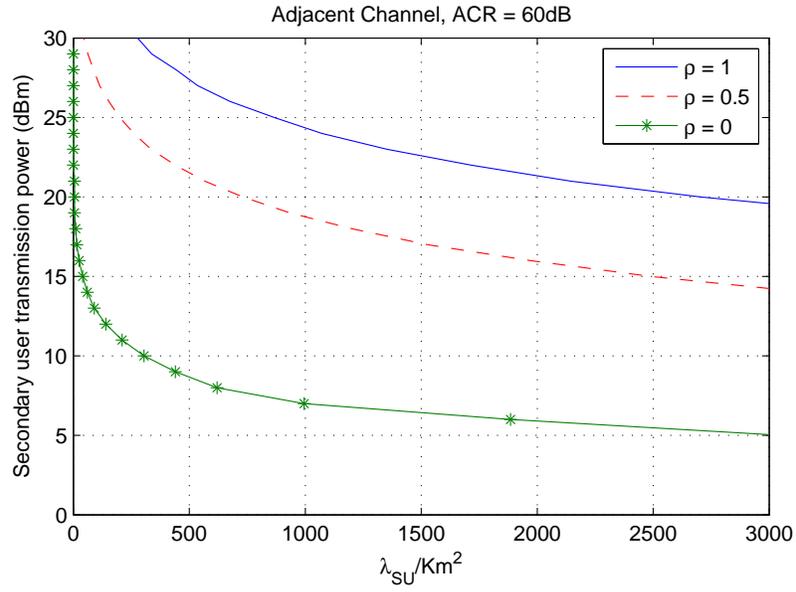}\\
  \centering \caption{Impact of different correlation coefficients ($\rho$) on the feasibility of the secondary access when $Pr(\tilde{\xi}_{j}\leq I_{thr})\geq 90\%$ at $r_{j}=5$km, the primary receiver is the DME ground transponder}
  \label{fig:MaxSUdensity}
\end{figure}

The individual interference threshold $I_{thr}$ required to satisfy \eqref{eq:ProtectionDME} is given in Fig. \ref{fig:IthrDiffCorrelation}. It shows the impact of $\rho$ on the required $I_{thr}$ for accessing a co-channel and an adjacent channel with ACR of 60dB. We observe that the margins to cope with the uncertainty for different values of $\rho$ does not change much when the density of secondary users per km$^2$ ($\lambda_{SU}$) increases. However, the uncertainty margin significantly varies for co-channel and adjacent use cases, i.e. for different $A_{thr}$ values. Considering that secondary users transmit in an adjacent channel with ACR of 60dB, it is observed in Fig.~\ref{fig:MaxSUdensity} that the impact of fading uncertainty is critical for high-power secondary users (above 10dBm). However, the operation of a dense secondary network for indoor coverage is feasible even if the secondary users cannot accurately estimate the propagation loss. 

\begin{figure}[h]
  \centering \includegraphics[width=.68\textwidth]{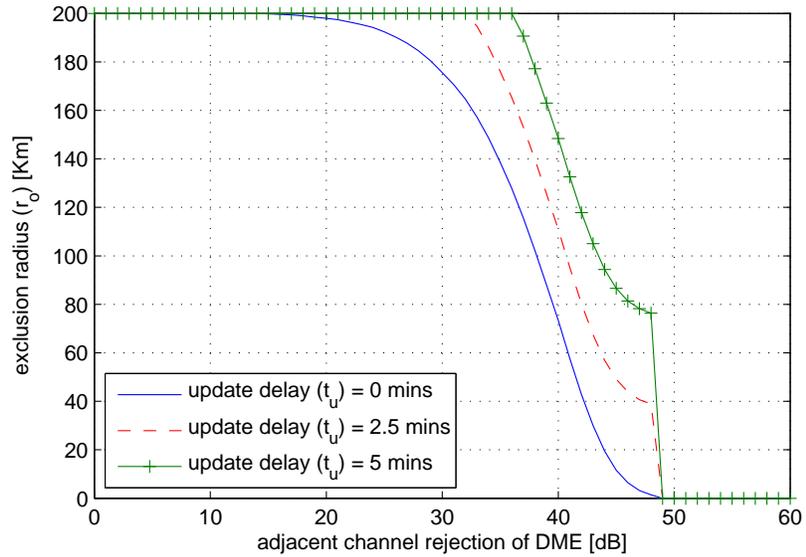}\\
  \centering \caption{Exclusion radius as a function of adjacent channel rejection of DME for different update delays when $\lambda_{SU}=20/$km$^2$, the primary receiver is the DME airborne interrogator}
  \label{fig:AirborneRoVsDensity}
\end{figure}

\begin{figure}[h]
  \centering \includegraphics[width=.68\textwidth]{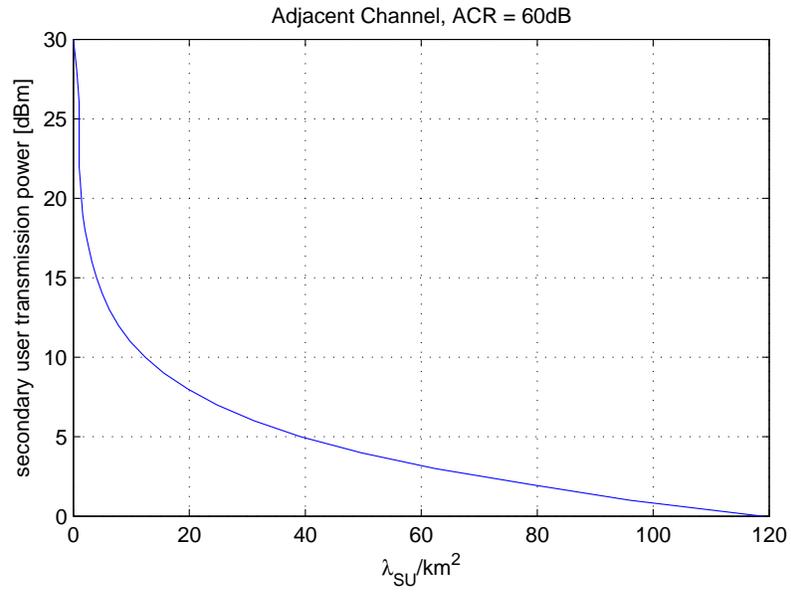}\\
  \centering \caption{Maximum secondary user transmission power for a given $\lambda_{SU}$ when no exclusion region is needed ($r_o = 0$km), the primary receiver is the DME airborne interrogator}
  \label{fig:AirPowerVsDensity}
\end{figure}

Now, let us consider the airborne interrogator as the primary victim. Recall that the update delay of 5 minutes can lead to the error region of 75 km radius, which is almost equivalent to not having the database. The exclusion region needed to satisfy \eqref{eq:ProtectionDME} is shown in Fig.~\ref{fig:AirborneRoVsDensity}. The impact of the update delay is significant only when ACR is lower than 50dB. When ACR is 60dB, no exclusion region is required even if long update delay is experienced in the communication between the secondary user and the real-time database. Fig.~\ref{fig:AirPowerVsDensity} shows the combination of secondary users density and transmission power that do not require fast database update, i.e. no requirement for exclusion region. The figure indicates that dense secondary network accessing adjacent channels is feasible when the transmission power is about 0dBm even if no information on the location of the primary victim is provided.

\section{Conclusions and Future Work}
\label{sec:Conclusions}
We analyzed the feasibility of large scale indoor broadband secondary access to the 960-1215 MHz spectrum when uncertainties on the fading and the location of the primary receiver are present. Cumulant-based approximations have been employed to derive the probability distribution of the aggregate interference in the presence of uncertainties. The main contributions in this paper are twofold:
\begin{itemize}
\item We proposed a practical secondary sharing scheme considering the characteristics of different primary receivers (DME ground transponder and airborne interrogator). Then, we identified uncertainties in the estimation of propagation loss incurring from the proposed sharing scheme.
\item The feasibility of large scale secondary access has been evaluated in terms of the number of secondary users which are able to operate with an acceptable transmission probability and the exclusion region size imposed on the secondary users.
\end{itemize}

We conclude that massive indoor secondary access to adjacent channels (ACR~$\geq~60$dB) is feasible even if secondary users are not capable of accurately estimating the propagation loss nor have accurate knowledge of the location of the airborne interrogator. Numerical results show that dense secondary users ($\lambda_{SU}>1000/\textrm{km}^2$ for ground transponder and $\lambda_{SU}>100/\textrm{km}^2$ for airborne interrogator) can have access to adjacent channels with a high transmission probability ($\geq~90\%$) or small exclusion region size. 

Since the indoor secondary use of 960-1215 MHz spectrum is identified feasible, the capacity analysis of the secondary system taking self-interference and power control into account remains as an interesting future work. Location-dependent availability of the secondary access and its economic value are also to be investigated.

\section*{Acknowledgment}
The research leading to these results has received funding from the European Union's Seventh Framework Programme FP7/2007-2013 under grant agreement n$^o$~248303 (QUASAR). 

\bibliographystyle{IEEEtran}
\bibliography{uncertaintyTVT_KWS}

\begin{table}[ht]
\renewcommand{\arraystretch}{1.1}
\centering
\caption{Parameters used for numerical experiments}\label{tab:param}
  \begin{tabular}{l|l}
    \hline\hline
    \multicolumn{2}{c}{\textbf{Parameters for ground transponder}}\\
    \hline
    primary user transmission power & 60dBm/MHz\\
    \hline
    path loss constant ($C$) & $4.5 \times 10^{-13}$\\
    \hline
    path loss exponent ($\alpha$) & 3.5\\
    \hline
    Fading standard deviation ($\sigma_{X_j}^{dB}$) & 10dB\\
    \hline
    height of the transponder & 30 m\\
    \hline\hline
    \multicolumn{2}{c}{\textbf{Parameters for airborne interrogator}}\\
    \hline
    primary user transmission power & 55dBm/MHz\\
    \hline
    path loss constant ($C$) & $5.7 \times 10^{-10}$\\
    \hline
    path loss exponent ($\alpha$) & 2.0\\
    \hline
    height of the interrogator ($h$) & 1 km\\
    \hline\hline
    \multicolumn{2}{c}{\textbf{Common parameters}}\\
    \hline
    radius of interference aggregation ($R$) & 200 km\\
    \hline
    building penetration loss & 10dB\\
    \hline
    DME antenna gain & 5.4dBi\\
    \hline
    secondary user antenna gain & 0dBi\\
    \hline
    secondary user transmission power & 1dBm/MHz\\
    \hline
    secondary user height & 1.5 m\\
    \hline
  \end{tabular}
\end{table}


\end{document}